\documentclass[12pt,preprint]{aastex}

\begin{document}

\title{The Extended X-ray Nebula of PSR J1420--6048}

\author{Adam Van Etten\altaffilmark{1}, Roger W. Romani\altaffilmark{1}}

\altaffiltext{1}{Department of Physics, Stanford University, Stanford, CA 94305}


\begin{abstract}

The vicinity of the unidentified EGRET source 3EG J1420--6038 has undergone extensive study in the
search for counterparts, revealing the energetic young
pulsar PSR J1420-6048 and its surrounding wind nebula
as a likely candidate for at least part of the emission from this bright and extended gamma-ray source.
We report on new Suzaku observations of PSR J1420--6048, along with analysis of
archival XMM Newton data.  The low background of Suzaku permits mapping of the extended
X-ray nebula, indicating a tail stretching  
$\sim 8 \arcmin$ north of the pulsar.  The X-ray data, along with archival radio and VHE data,
hint at a pulsar birthsite to the North, and yield insights into its evolution 
and the properties of the ambient medium.  We further explore such properties by modeling
the spectral energy distribution (SED) of the extended nebula.

\end{abstract}

\section{Introduction}
The campaign to identify 3EG J1420--6038 has revealed
sources across the electromagnetic spectrum from radio to VHE $\gamma$-rays.
The complex of compact and extended radio sources in this region is referred to as the 
Kookaburra \citep{robertsetal99}, and covers nearly a square degree along the Galactic plane.
Within a Northeasterly excess in this complex labeled ``K3'' 
\citet{d'amicoetal01} discovered PSR J1420--6048 (hereafter J1420), a young energetic pulsar 
with period 68 ms, characteristic age $\rm \tau_c = 13$ kyr, and spin down energy 
$\rm \dot E = 1.0 \times 10^{37} \, erg \, s^{-1}$.  The NE2001 dispersion measure model \citep{candl02}
of this pulsar places it 5.6 kpc distant. 
Subsequent ASCA observations by \citet{robertsetal01b} revealed extended X-ray emission around this pulsar, 
and \citet{ngetal05} further examined the K3 pulsar wind nebula (PWN) with Chandra
and XMM-Newton, resolving a bright inner nebula along with fainter emission extending
$\sim 2\arcmin$ from the pulsar.
\citet{aharonianetal06b} report on the discovery of two bright VHE $\gamma$-ray sources coincident 
with the Kookaburra complex.  HESS J1420-607 is centered just north of J1420, with best fit extension 
overlapping the pulsar position.  The other H.E.S.S.\ source appears to correspond to the Rabbit nebula
half a degree southwest, which is also 
observed in the radio  \citep{robertsetal99} and X-ray \citep{robertsetal01a}.
Most recently, PSR J1420-6048 was detected by the Fermi Large Area Telescope (LAT) \citep{abdoetal09}.
This crowded region clearly merits further study, and we report on new X-ray results obtained with 
Suzaku and XMM-Newton, as well as SED modeling of the K3 nebula.

\section{Data Analysis}
The Suzaku pointing (obsID 503110010) occurred on January 11-12 2009 for a total of 50.3 ks.  
We utilize the standard pipeline screened events, and analyze the XIS
front side (XIS0 and XIS3) and back side (XIS1) illuminated chips with XSelect version 2.4.  
We also obtained recent archival XMM data to augment the Suzaku data; observation 0505840101 occurred on 
February 15 2008, for 35.0 ks, while observation 0505840201 added 5.6 ks.  
The second data set has a slightly different CCD placement, and suffers from high background, so we only 
use the 35.0 ks of data.  We apply the standard data processing, utilizing SAS version 9.0. After  
screening the data for periods of high background 19.9 ks remain with the MOS chips. The PN chip suffers greatly from flaring, and we discard this data.  Spectral fits are 
accomplished with XSPEC version 12.5.

\subsection{Broadband Morphology and Point Sources}
Suzaku X-ray emission is peaked in the vicinity of the pulsar, with a bright halo extending $\sim3 \arcmin$
and a fainter tail extending north $\sim 8 \arcmin$.  A number of other excesses of emission 
correspond to point sources, as discussed below.  
Figure 1 shows the Suzaku data in the 2--10 keV band, which highlights the 
extended PWN emission to the north.  Also depicted is the XMM exposure, clearly showing
a number of point sources, though no obvious extended emission is apparent.

\begin{figure}[h!]
\plottwo{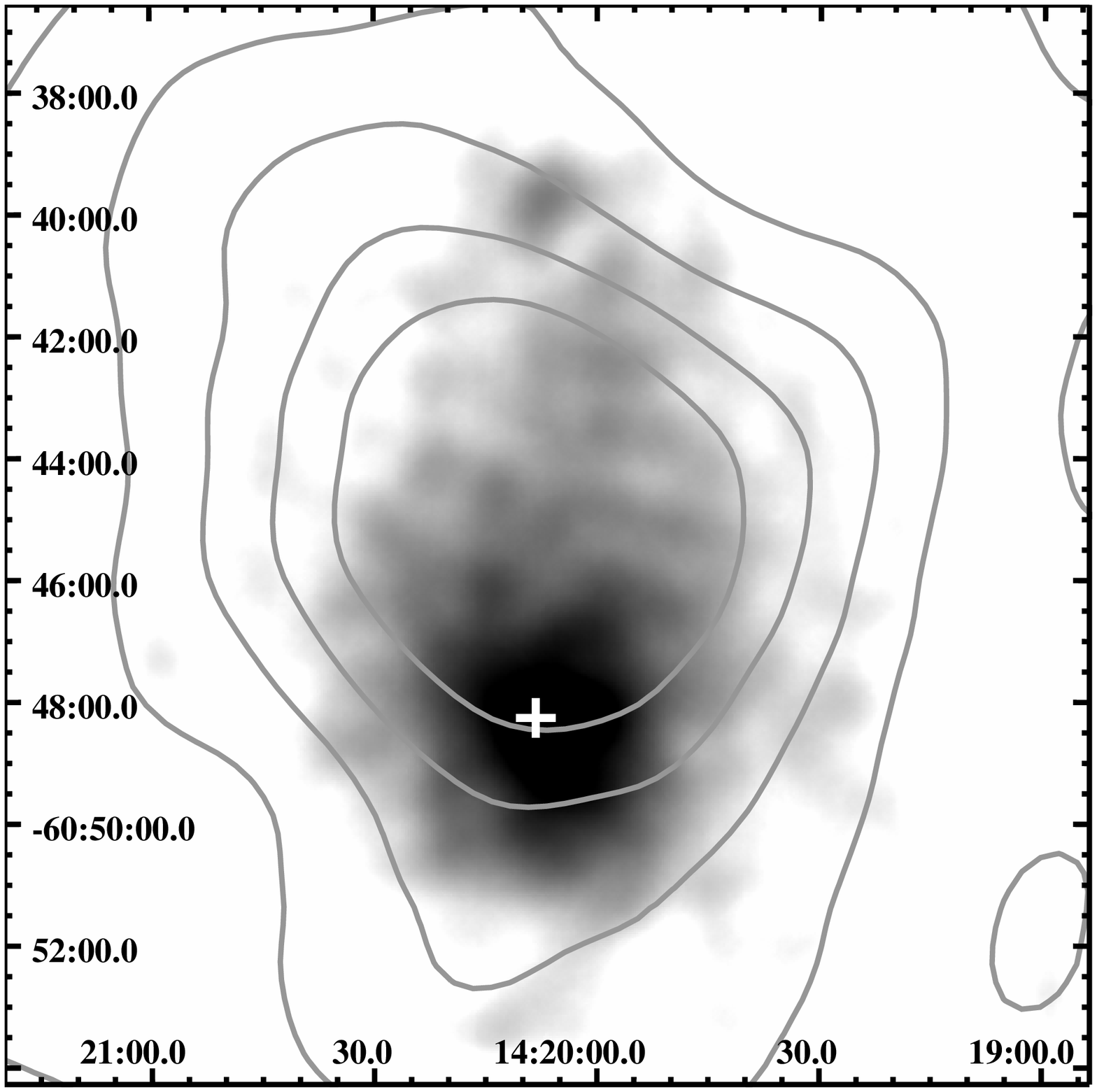}{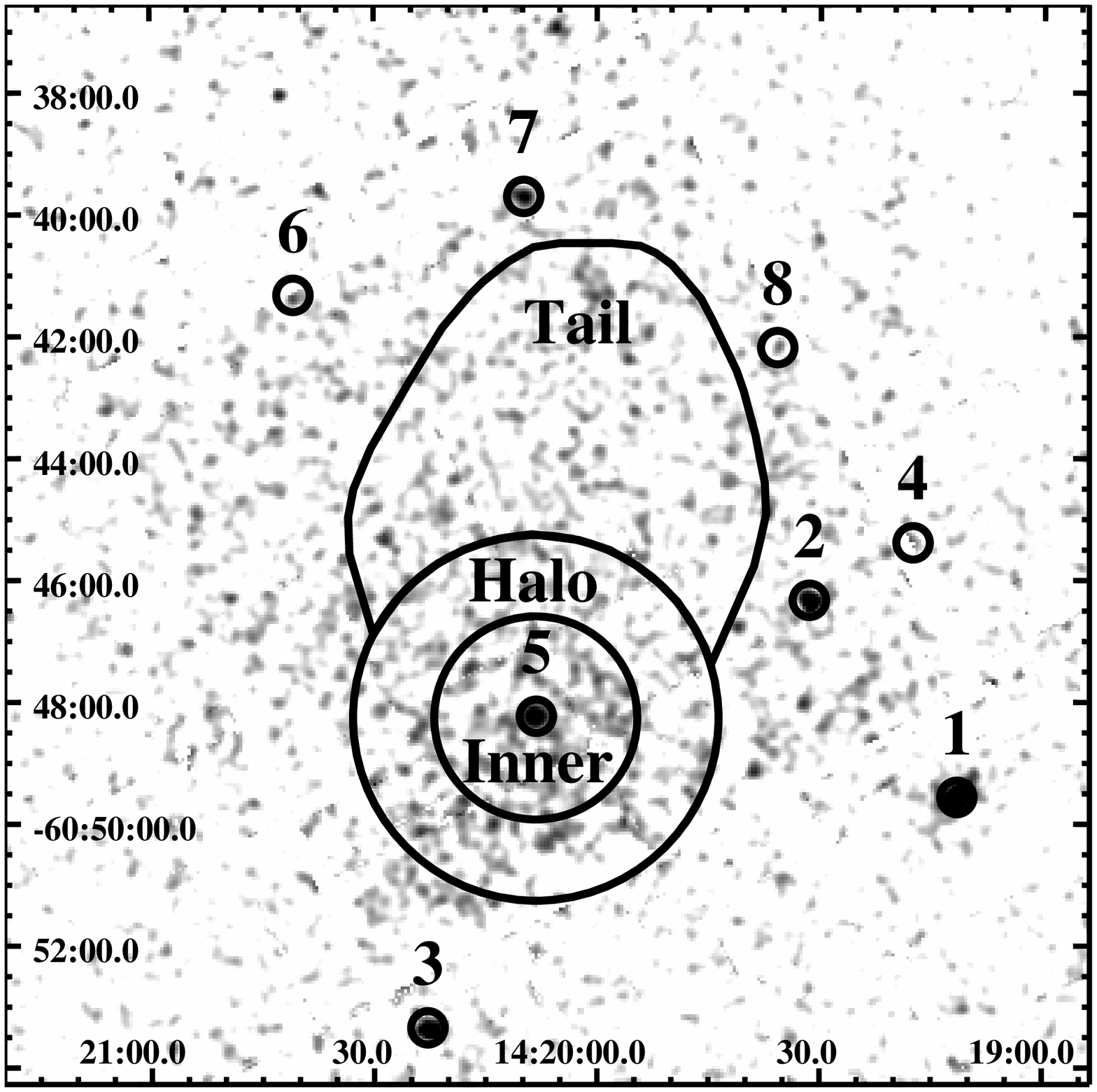}
\caption{
Left: Suzaku XIS0+XIS1+XIS3 2-10 keV image overlaid with H.E.S.S contours and
smoothed with a $40\arcsec$ Gaussian. PSR J1420--6048 is indicated by
the white cross.
Right: XMM EPIC 0.2-12 keV mosaic showing named spectral extraction regions
and numbered point sources, smoothed with a $10\arcsec$ Gaussian; no extended 
tail emission is apparent. 
}
\end{figure}

To identify X-ray point sources we use the SAS source detection
function edetect$\_$chain on the XMM MOS chips and search in
two energy bands of 0.3--2 keV and 2--10 keV for sources with a probability $P < 10^{-13}$
of arising from random Poissonian fluctuations.  The source
detection algorithm also attempts to determine source extension via a
Gaussian model, though all detections are consistent with a point source. Counts 
are therefore extracted from a 15 pixel ($16.5 \arcsec$) radius circle.
A total of 8 sources pass this test, 4 of which also appear
 in \citet{ngetal05}: PSR J142--6048 (source 5 in our dataset), the X-ray sources denoted star 1 (source 1) and star 3
(source 2), and another point source to the southeast (source 3) unlabeled by \citep{ngetal05}
 but visible in their XMM exposure.  
This source to the southeast is also a field star, as it appears quite bright in DSS2 red images.  
Of the four remaining sources, only one, a hard bright source $8.5 \arcmin$ north of the J1420
labeled source 7, lacks an optical counterpart.  Source 7 also overlaps a radio hotspot to the north.  
Below we list the properties of these sources, defining
the hardness ratio as: 
HR=$ \rm (C_{hi}-C_{lo})/(C_{hi} + C_{lo})$ where $\rm C_{lo}$ and 
$\rm C_{hi}$ are MOS counts in the 0.3--2 keV and 2--10 keV bands,
respectively.  
It is worth noting that PSR J1420--6048 is only the fifth brightest point source in the XMM field and 
that all 8 XMM sources appear as excesses in the soft band Suzaku data as well.  
All point sources are quite soft, save for J1420 and source 7. 
         
\begin{table}[!h]
\caption{XMM Source Properties}
\begin{tabular}{lccccc}
No. ($\tablenotemark{*}$) & R.A. & Dec. & Pos. Err.$\arcsec$  & Counts & HR   \\
\hline
1 (Star 1)	&  $14:19:11.52$  &  $-60:49:34.00$ &  $0.26$ &  $462 \pm 27$ &  $-0.99 \pm  0.026$	\\   
2 (Star 3)	&  $14:19:31.48$  &  $-60:46:20.29$ &  $0.39$ &  $146 \pm 18$ &  $-0.85 \pm  0.12 $	\\ 
3 (Unlabeled)	&  $14:20:22.72$  &  $-60:53:21.47$ &  $0.45$ &  $164 \pm 19$ &  $-1.00 \pm  0.069$	\\   
4 		& $14:19:17.61$  &  $-60:45:23.45$ &  $0.61$ &  $123 \pm 17$ &  $-0.81 \pm  0.12 $	\\  
5 (PSR J1420--6048)&  $14:20:08.19$  &  $-60:48:14.85$ &  $0.63$ &  $150 \pm 20$ &  $ 0.95 \pm  0.072$	\\   
6 		&  $14:20:40.75$  &  $-60:41:20.22$ &  $0.79$ &  $40  \pm 11$ &  $-0.79 \pm  0.30 $	\\  
7 		&  $14:20:09.78$  &  $-60:39:42.86$ &  $0.85$ &  $62  \pm 14$ &  $ 0.79 \pm  0.18 $	\\  
8 		&  $14:19:35.85$  &  $-60:42:11.39$ &  $1.16$ &  $34  \pm 11$ &  $-0.68 \pm  0.36 $	\\  

\hline
\end{tabular}
\tablenotetext{*}{\citep{ngetal05} counterpart}
\label{srcprop}
\end{table}

On a larger scale, extended emission is observed in all wavebands.
Australia Compact Telescope Array (ATCA) observations within the error ellipse 
of 3EG J1420--6038 (which is broad enough to encompass both the K3 wing and the Rabbit nebula)
by \citet{robertsetal99} revealed 
the ``K3'' excess, a resolved knot of emission surrounding the pulsar of
flux density 20 mJy at 20 cm with index $\alpha = -0.4 \pm 0.5$.  Adjacent 
is the ``K2 wing,'' with 1 Jy at 20 cm and index of $-0.2 \pm 0.2$.  
Closer inspection of the both the 13 cm and 20 cm continuum maps reveal that J1420 lies on the southeastern rim
of an apparent radio shell $\approx 3\arcmin$ in radius.  This shell is also apparent in the SUMSS 843 MHz 
map of the region.  
The center of this shell coincides with a dearth of emission in Spitzer 
$8 \, \mu \rm m$ maps.   
We place extended PWN upper limits in the radio by measuring
 the 843 MHz, 20 cm and 13 cm flux densities from the entire shell 
(which is significantly larger than the X-ray extension), finding
0.61, 0.75 Janksy and 0.58 Jansky, respectively. 
We deem these flux densities to be upper limits since the poor spatial 
resolution of Suzaku prevents ascertaining how the radio and X-ray 
emitting regions relate.  
We also remeasure the K3 excess and find flux densities of 
15 mJy, 19 mJy and 17 mJy at 843 MHz, 20 cm and 13 cm, respectively, 
consistent with the result of \citet{robertsetal99}).
At higher energies HESS J1420--607 
shines at 13\% 
of the Crab \citep{aharonianetal06b}, with photon index of 2.2 and extent of $3.3\arcmin$ 
centered $2.6\arcmin$ north of the pulsar.
The H.E.S.S.\ spectrum is extracted from a $9.6\arcmin$ circle to minimize contamination from HESS J1418-609
(The Rabbit) $33\arcmin$ to the southwest. 
          
\begin{figure}[h!]
\plottwo{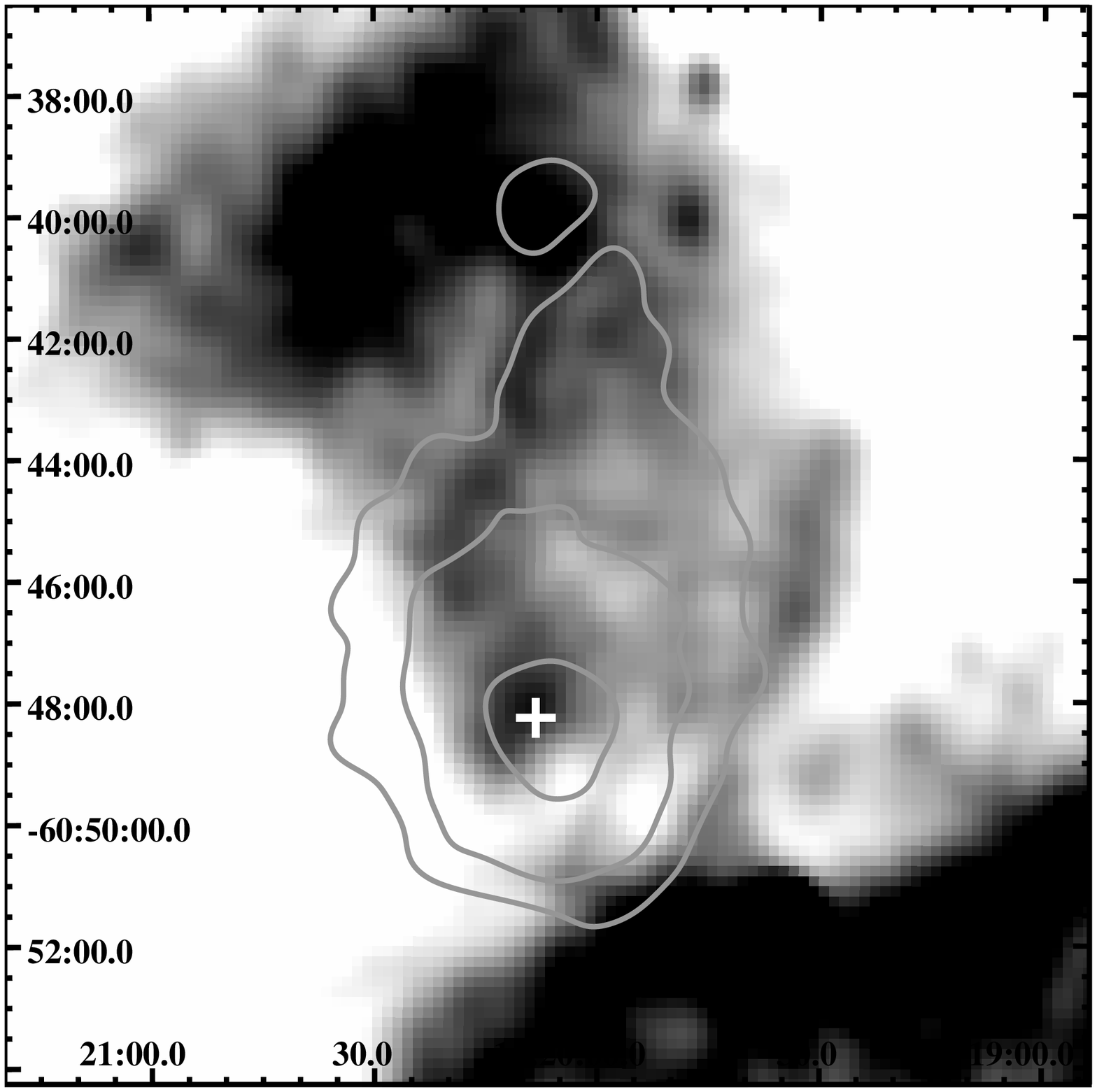}{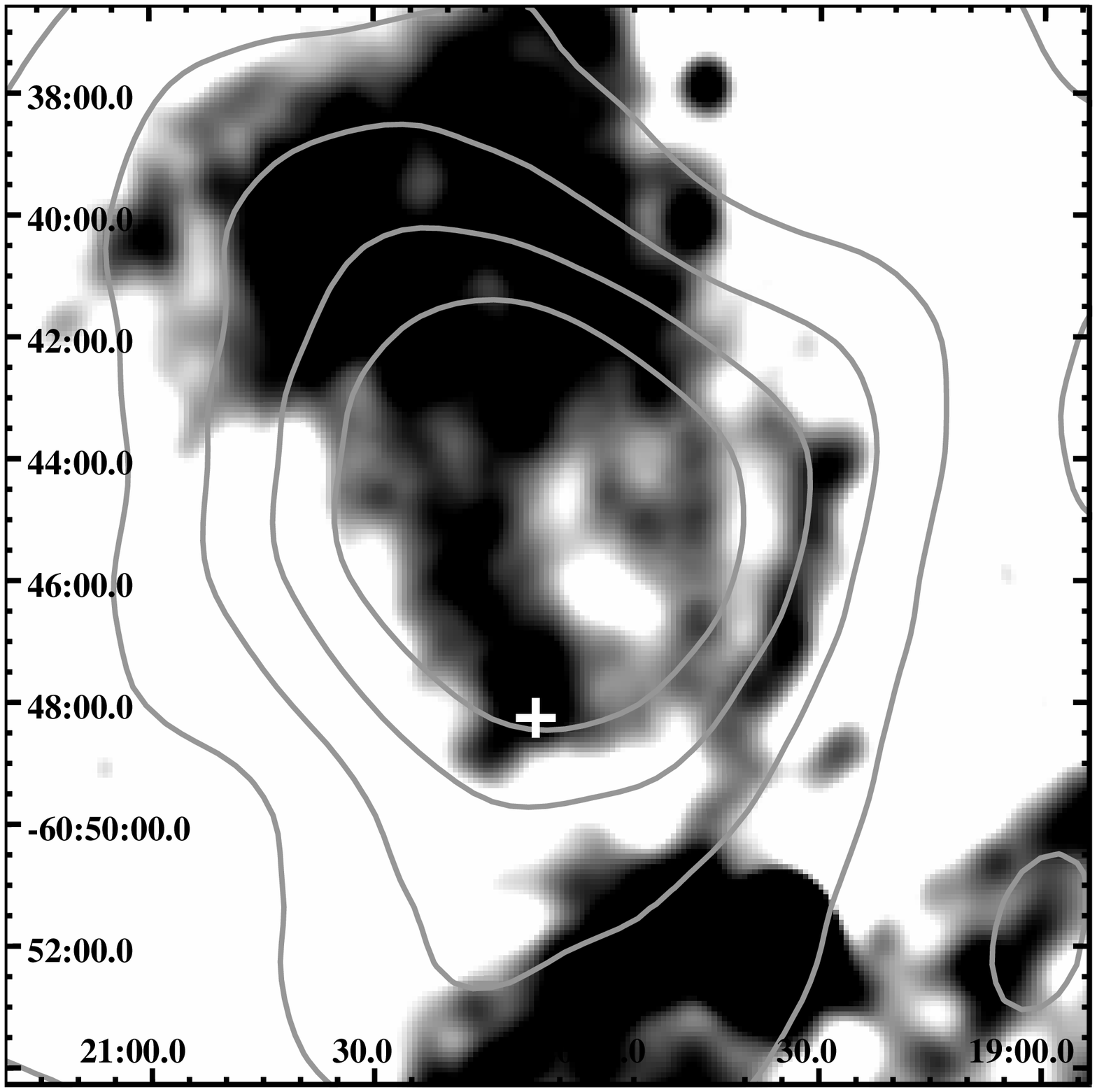}
\caption{
Left: 20 cm ATCA image showing the
possible cavity
northwest of PSR J1420--6048 (indicated by a white cross) after \citet{robertsetal99}.
Suzaku contours are overlaid, with the northern blob due to source 7.
Right: SUMMS 843 MHz image with H.E.S.S. contours overlaid.  
The bright emission to the south is the HII region G313.5+0.2. 
}
\end{figure}
   
\subsection{Spectrum}
The XMM point source identification described above is valuable since the broad Suzaku PSF hinders the 
disentanglement of point sources from extended emission.  We are accordingly free to define extended regions for spectral 
analysis while steering clear of the X-ray point sources; we label these ``Inner,'' ``Halo,'' and ``Tail''.  
All spectral extraction regions lie external to the Suzaku $1\arcmin$ half power radius of the 8 point sources (except
the very faint and soft source 8, which slightly overlaps the ``Tail'').   
The Northernmost emission is encompassed by the ``Tail'' region 
which extends north of the pulsar from $3\arcmin \sim 8\arcmin$.  
Suzaku does best with large regions, and accordingly we capture the pulsar and surrounding PWN emission with a
circular region of radius $3\arcmin$ (``Halo'').  To best isolate the pulsar
and inner PWN we also use a $1.7\arcmin$ circle we call ``Inner,''
which is the minimum size recommended for Suzaku spectral analysis.
The Suzaku background region occupies 
the northern and eastern edges of the field of view.  

We extract Suzaku XIS spectra with XSelect from all three active XIS chips, while   
response files (both ARF and RMF) 
are generated with the xisresp script.  The ASCA FTOOL addascaspec is utilized to combine 
the spectra and responses of the XIS front side illuminated chips (XIS0 and XIS3).  
The XIS1 back side illuminated chip possesses a markedly different response function from the 
front side illuminated chips, and so is analyzed in parallel rather than added to the other two chips.  
Finally, spectra are binned to a minimum of 
30 counts per bin. The high XMM background prevents an adequate fit to the extended emission 
(flux errors of $\sim100$\%). 
 We are able, however, to extract the spectrum from the ``Inner'' region encompassing the pulsar.
To minimize calibration errors due to differing chips and distance from the pointing axis, 
we select an XMM background region 
on the northeast corner of the 
central chip containing PSR J1420-6048.   We extract XMM spectral data with the SAS
function evselect, create responses with the functions rmfgen and arfgen,
and group to 30 counts per bin.  

To best probe spectral index variations we make a global estimate for the best fit hydrogen column density.  
To maximize the 
signal to noise, we simultaneously fit $n_H$ 
for the Suzaku ``Halo'' spectrum and the ``Inner'' XMM spectrum.  For an absorbed power law model
we find a 
$n_H$ of $4.1_{-0.4}^{+0.6} \times 10^{22} \rm \, cm^{-2}$ (90 \%
single parameter error).
An individual fit to the inner PWN region indicates that an absorbed power law model provides 
the best fit to the data; addition of a blackbody, neutron star atmosphere, 
or thermal plasma (mekal) component does not improve the fit.  
Below 2 keV the fit to all regions significantly underestimates the flux.  This feature
is not well fit with a thermal plasma, neutron star atmosphere, or blackbody, and might simply
constitute excess soft emission from the myriad faint sources in this crowded region.
The $4.1_{-0.4}^{+0.6} \times 10^{22} \rm \, cm^{-2}$ column 
is marginally consistent with the value measured 
by \citet{robertsetal01b} of $2.2 \pm 0.7 \times 10^{22} \rm \, cm^{-2}$ 
using ASCA GIS data ($3\arcmin$ radius), and
matches well with the short Chandra ACIS exposure fit by \citet{ngetal05}
($2\arcmin$ radius aperture) of $5.4_{-1.7}^{+2.2} \times 10^{22} \rm \, cm^{-2}$
(1$\sigma$ error).  The nominal total Galactic H column in this direction is estimated as 
$1.6 \times 10^{22} \rm \, cm^{-2}$ \citep{kalberlaetal05} and $2.1 \times 10^{22} \rm \, cm^{-2}$
 \citep{dandl90}.

With $n_H$ fixed at $4.1 \times 10^{22} \rm \, cm^{-2}$, we extract spectra from the three regions described above.  
Table 1 lists the results of fitting an absorbed power law to these regions in the 2-10 keV range; errors are 
90\%
 single parameter values.  
We quote 2--10 keV fluxes to compare to previous work, minimize sensitivity to 
the hydrogen column density fit, and mitigate soft X-ray contamination in this crowded region.  
In the inner $100\arcsec$ region
we fit simultaneously the XMM and Suzaku data, and find a power law index of $1.8 \pm 0.1$.  
The entire $3 \arcmin$  
``Halo'' is fit with an index of $2.0 \pm 0.1$, 
implying a steeper power law near the edge of the nebula.  To further 
explore this, we extract the spectrum from an annulus of radius $1.7\arcmin - 3\arcmin$ from J1420.  
For such annular regions
the Suzaku response normalization proves unreliable, though the spectral shape is still of value.  
We find a power law index of $1.9 \pm 0.1$; 
this is indeed softer (though within errors) of the inner PWN.  Further spectral 
softening is hinted at in the outer nebula ``Tail,'' which boasts an index of $2.1 \pm 0.1$.

\begin{table}[!h]
\caption{Spectral Fits}
\begin{tabular}{lccccc}
Region & $N_{\rm H}$ & $\Gamma$& abs. flux& unabs. flux & $\chi^2$/dof \\
& $\times10^{22}$cm$^{-2}$& & $f_{2-10}$\tablenotemark{\dagger}&$f_{2-10}$\tablenotemark{\dagger}& \\
\hline
Inner$\tablenotemark{\Diamond}$ & $4.1$\tablenotemark{*} & $1.82 \pm 0.13$ & $1.19_{-0.20}^{+0.24}$ & $1.62_{-0.28}^{+0.32}$ & 333/345 \\
Halo & $4.1$\tablenotemark{*} & $2.00\pm0.10$ & $2.45_{-0.31}^{+0.35}$ & $3.40_{-0.43}^{+0.48}$ & 219/309 \\
Tail & $4.1$\tablenotemark{*} & $2.14\pm0.16$ & $1.26_{-0.25}^{+0.30}$ & $1.79_{-0.35}^{+0.42}$ & 138/224 \\
\hline

\end{tabular}

\tablenotetext{\Diamond} {Simultaneous fit to Suzaku and XMM data}
\tablenotetext{*}{held fixed}
\tablenotetext{\dagger}{2-10\,keV fluxes in units of $10^{-12}{\rm erg\,cm^{-2}\,s^{-1}}$}
\label{pwnspec}
\end{table}

These spectral values are largely consistent (though with smaller errors) with previous data fits.
\citet{ngetal05} extract the spectrum of the inner nebula 
from a 2' circle, which gives (for $n_H$ fixed at $5.4 \times 10^{22} \rm \, cm^{-2}$) $\Gamma = 2.3_{-0.8}^{+0.9}$ 
(projected multidimensional 1 sigma error), unabsorbed 2-10 keV flux 
$1.3 \pm0.14 \times 10^{-12} \rm \,erg\,cm^{-2}\,s^{-1}$ (1 sigma single parameter error).  
\citet{robertsetal01b} measure a power law index of $1.6\pm0.4$ and a 2-10 keV flux of 
$4.7 \times 10^{-12} \rm \,erg\,cm^{-2}\,s^{-1}$ for an extraction region of size $3\arcmin$.

\section{SED Fitting}

Matching SED data points can help constrain physical properties of the source. 
To this end we we apply a two-zone time-dependent
numerical model with constant injection luminosity. 
We inject a power law spectrum of relativistic
particles (either electrons or protons) with a high energy exponential cutoff into zone 1, 
and then evolve this spectrum over time according to radiation losses from synchrotron and
IC (Klein-Nishina effects included).  The resultant spectrum is subsequently injected into zone 2 
and further evolved.
Adiabatic cooling is ignored, given that offsetting heating effects of the SNR
reverse shock interaction with the PWN may significantly alter particle energetics.
The typical timescale for the reverse shock to reach the PWN is $\sim7$ kyr \citep{randc84},
while the characteristic age of the pulsar is only 13 kyr, so this perturbation may still be felt. 
The complex nature of this interaction is also why we use a constant injection luminosity
rather than allowing injection luminosity to vary with pulsar spin down power.
Injection (and evolution) occurs in time steps much smaller than the assumed age.
We consider a low density environment with density $n$ = 1 cm$^{-3}$
and photon fields comprised of CMBR, 
near IR,
and far IR (though starlight is unimportant due to Klein-Nishina suppression of IC at 
high photon energies).
Our two zones are defined as the inner $100\arcsec$ region
surrounding the pulsar (zone 1) and the broader extended nebula (zone 2).  
Figures 3 and 4 show the SED of J1420.  While we must select spectral 
extraction regions such that interference from stellar sources is minimized
and region surface brightness remains significantly above background, the 
nebular flux extends beyond these regions.  An estimate of the total
nebular flux is essential if we wish to
compare the X-ray flux with the VHE flux, which is extracted
from a large $\approx 10 \arcmin$ circle.  
To estimate the total extended nebula flux,
we extract 2-10 keV counts from a large 
ellipse $5 \arcmin \times 6 \arcmin$ in radius which seems to encompass
the majority of the X-ray flux (excluding point
sources within this field) and find 40\%
more counts in this region than the combined ``Halo'' plus ``Tail'' minus ``Inner'' region.
We therefore extrapolate that the total extended nebular flux is 40\%
greater than the combined ``Halo'' plus ``Tail'' minus ``Inner'' fluxes.
Accordingly, for SED modeling of the extended nebula 
we plot the spectral points from the ``Tail'' region, scaled up 
to represent this estimated 2-10 keV flux of 
$5.0 \times 10^{-12} \, \rm erg \, cm^{-2} \, s^{-1}$.

wn Figures 3 and 4 We plot the gamma-ray
flux from HESS J1420--607, along with radio and X-ray data.  
Also indicated are the Fermi 
LAT 
one-year and ten-year $5\sigma$ flux limit for a point source
residing in a background $10\times$ greater than the extragalactic 
background, which is reasonable for the Galactic plane.

\subsubsection{Leptonic gamma-ray emission model}

For an X-ray spectral index 
approximately flat on the SED plot one cannot 
easily match the X-ray data to the H.E.S.S. data.  A single electron injection component
works, but only after significant fine tuning of injection and ambient medium parameters.  
A dual component injection comprises the other alternative, though 
the attentive reader might note that the northern X-ray nebula spatially corresponds quite
well to the H.E.S.S. position and size, hinting at a single electron population responsible for 
both types of emission.  If one assumes all particles evolve in the same spatial region 
(and hence same magnetic field and ambient photon field) one cannot independently fit the 
X-ray data to one component and VHE data to another significantly different component. 
Therefore, we refrain from displaying the unsatisfactory dual component fits, and focus on a 
single component model.   

A single electron component demands that the H.E.S.S. detection
arise from IC scattering of far IR photons, as the CMB provides insufficient seed photons to
account for the VHE peak given the constraints imposed by the X-ray data.
For the Galactic radius of J1420, 
far IR photons typically peak at 
$\approx T = 25$ K with a density $\approx 1$ eV$\rm \, cm^{-3}$
\citep{pms06}.  
We find a better fit with a denser far IR photon field (2 eV$\rm \, cm^{-3}$), 
which is reasonable given the large scatter
in ambient photon fields throughout the Galaxy. 
The starlight photon field is taken to have a temperature of 3500 K with density
1 eV$\rm \, cm^{-3}$.

The innermost $100\arcsec$ region is populated by young electrons
emitting in a relatively strong magnetic field.  
In the ideal case of a magnetic dipole, the pulsar surface magnetic field can be estimated as
$B = 3.2 \times 10^{19} \, (P \, \dot P)^{1/2} \, \rm G$
where $P$ is in seconds.  This gives a surface magnetic field of 
$2.4 \times 10^{12}$ G for PSR J1420--6048.  If one assumes the magnetic field is
dipolar out to the light cylinder, and then falls off as the inverse
of radius past the light cylinder, the volume averaged magnetic field 
out to $50 \arcsec$ from the pulsar is calculated as $11 \, \mu \rm G$.  Without
knowledge of where the bulk of the synchrotron emission stems from (given the 
poor spatial resolution of Suzaku) we
 therefore adopt a mean magnetic field of of $12 \, \mu \rm G$ for the innermost region.
Electrons responsible for the hard X-rays can diffuse out of this region
in as little as 400 years, as shall
be seen in Section 4.  We therefore evolve the inner zone 1 electrons over 500 years.
A power-law distribution of electrons with the standard index of 2 is injected, with 
a high energy cutoff at 400 TeV.  
For these parameters, we require
an energy in the innermost region of
$9 \times 10^{45} \, d_{5.6}^2$ erg 
in the form of leptons, where distance $d = d_{5.6}\times 5.6 \, {\rm kpc}$. 
The electron index of 2 underpredicts the K3 radio flux excess by a factor of $\sim4$.
A slightly softer electron index allows us to fit the radio points, but increases
the energy requirements by a factor of $\sim 3$ and results in an outer nebula radio
flux butting up against the upper limits established above.  
Better constraints on the non-thermal radio emission from this region would
aid greatly in differentiating between models.  

In a magnetic field of $12 \, \mu \rm G$ evolving over 500 years results in
a lepton spectral break at $\sim 100$ TeV.  This broken spectrum is
subsequently injected into zone 2, the outer nebula.
With this electron spectrum and for the photon field described above
the outer nebula requires a total lepton energy of 
$9 \times 10^{47} \, d_{5.6}^2$ erg 
in order to fit the H.E.S.S. data.  
This energy requirement
is reasonable for an extended PWN, and can be compared to the 
Vela X nebula possessing $\sim 10^{48}$ erg in the form of leptons \citep{dJ07}.
Evolving the electrons
over 12.5 kyr places the X-ray and VHE emitting electrons in the cooled regime for 
the $8 \, \mu$G field we select to match the X-ray data.  

\begin{figure}[h!]
\plotone{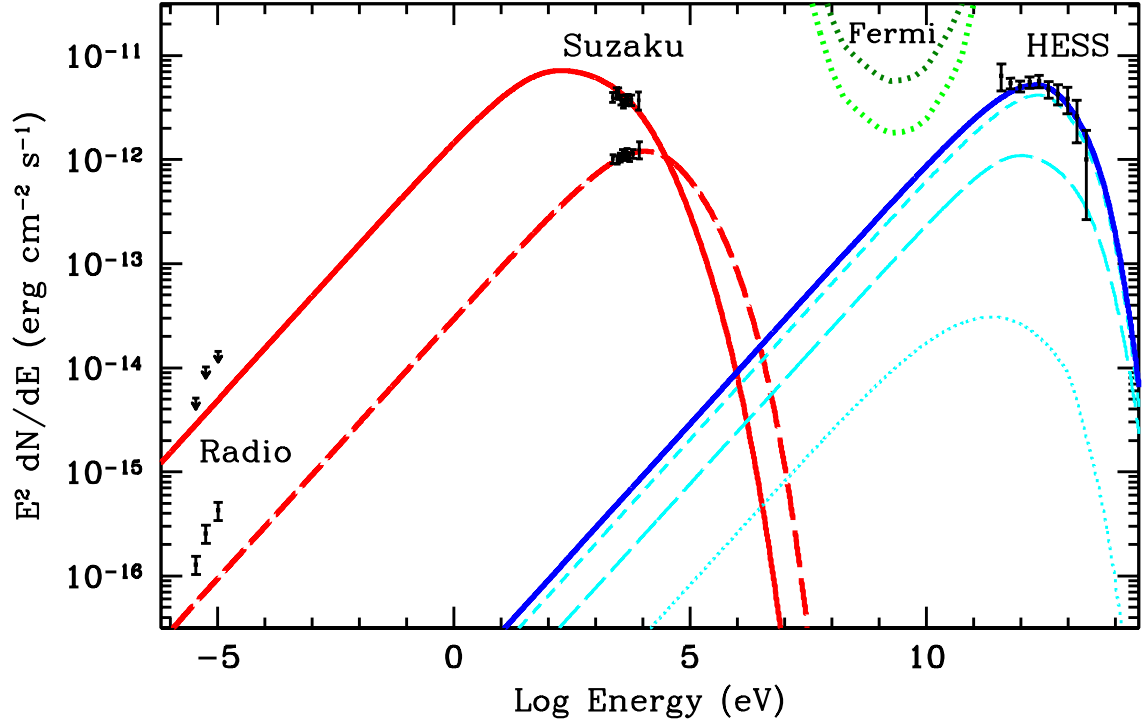}
\caption{
SED for single electron component
showing the radio, X-ray and VHE data.  
Solid lines indicate
the fit to the broad extended nebula, while the dot-dashed line
correspond to the inner PWN region.  Thinner lines denote the individual
IC components: starlight (dotted), CMB (long-dashed), and far IR (short-dashed).
Flux in the Fermi LAT realm is significantly below the 1 and 10-year detection lines.
}
\end{figure}

\subsubsection{Hybrid hadronic and leptonic gamma-ray emission model}

One can also model the VHE gamma-rays as the product of pion decay from proton proton interactions.
The timescale for pion production via p-p interactions is given by
$\rm \tau_{pp} \approx 1.5 \times 10^{8} \, (n/1 cm^{-3})^{-1}$ years 
\citep{b70}; this timescale is significantly greater than the expected age of the system, so the proton spectrum
is treated as static.   
For an injected power law proton spectrum, we calculate the photons from proton-proton interactions and 
subsequent $\pi^0$ and $\eta$-meson decay following \citet{ket06}. 
 Proton-proton interactions also yield $\pi^{\pm}$ mesons
which decay into secondary electrons, which we evolve over 13 kyr.  
IC and synchrotron
emission from the resultant secondary electron spectrum are subsequently calculated.

Secondary electrons evolved over 13 kyr cannot possibly account for the 
radio or X-ray data from the extended nebula for plausible magnetic fields.  We nevertheless 
indicate on Figure 4 the synchrotron radiation from secondary electrons. 
The X-ray data must therefore be attributed to synchrotron radiation from a primary electron population.  
We adopt the same two-zone lepton model described above, though with the outer nebula
parameters tweaked slightly.  We adopt the typical far IR field density of $1$ eV$\rm \, cm^{-3}$,
reduce lepton injection energy by $\sim 15$ \%
 in the outer nebula, and adopt a magnetic field of $9 \, \mu$G for this region. 
A magnetic field any greater than this is precluded both by the radio data and by increased
synchrotron cooling, which steepens the synchrotron slope in the Suzaku range.  
With these parameters gammas from pion
decay can account approximately equally with IC emission for the VHE data.  
The hadronic contribution to the H.E.S.S. detection can be modeled by a
a proton power law of index 2, cutoff at 200 TeV, and energy
$\rm E = 7\times10^{50} \, (n/1 \, cm^{-3}) \, d_{5.6}^2$ erg. 
The SED from this hybrid hadronic plus leptonic scenario is shown in Fig 4.
We omit the inconsequential IC radiation from secondary electrons.

\begin{figure}[h!]
\plotone{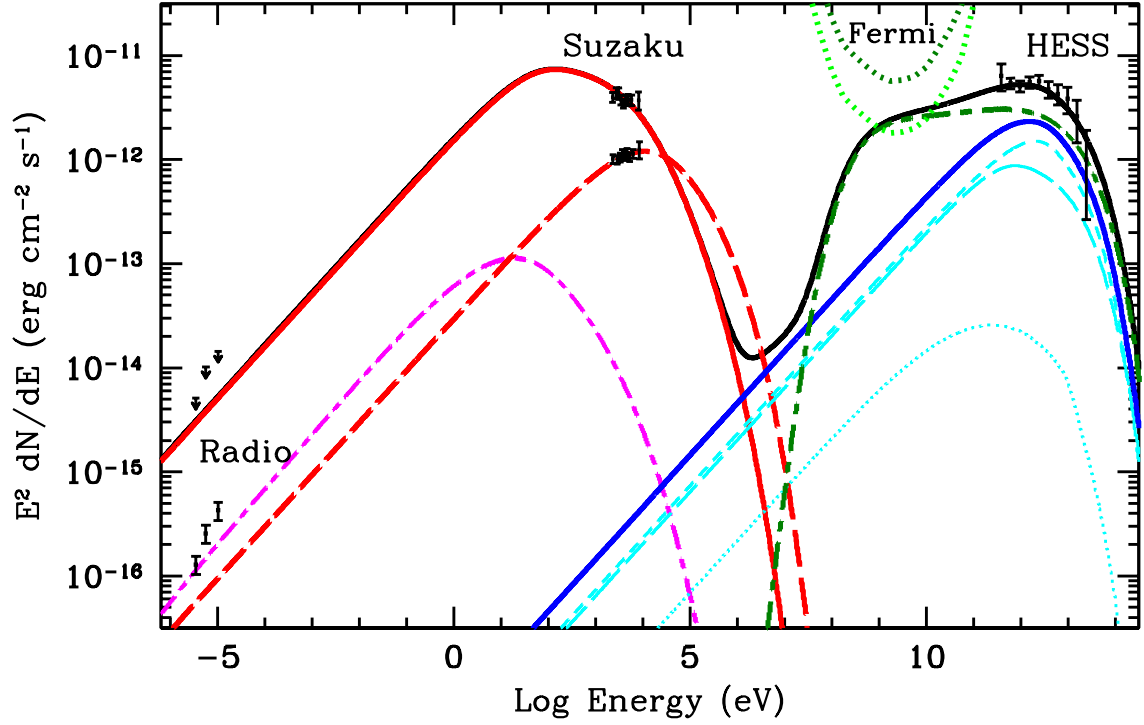}
\caption{
SED for hadronic plus leptonic model.  Dot-dashed lines indicate
pion decay and
synchrotron from secondary electrons
for a density of $n=1 \, \rm cm^{-3}$, solid lines
denote synchrotron and IC from the primary electron population. 
The long dashed line denotes synchrotron flux from the inner nebula, and
the solid line intersecting the H.E.S.S.\ points marks the total extended nebula flux.  
In this scenario Fermi detection could be expected after a few years.
}
\end{figure}

\section{Discussion}

X-ray flux extending $8\arcmin$ north of the pulsar indicates a spatial extent of 
$\rm R =  13 \, d_{5.6}$ pc.  Though this requires an average flow speed of under 1\%
the speed of light for a 13 kyr age, diffusion across magnetic field lines can be an extremely slow process.
For a cooling limited electron source with Bohm diffusion, the diffusion timescale varies with
size ($\theta$), magnetic field ($B$), and electron energy ($E_e$) as:
$$
t_{diff} \approx 68 \, (\theta/8\arcmin)^2 \,d_{5.6}^2 \, (B/10 \, \mu {\rm G}) \, (E_e/100 \, {\rm TeV})^{-1} \, \, {\rm kyr}
\eqno (1)
$$  
Therefore over the $\sim 10$ kyr lifetime of J1420,
the inferred presence of distant VHE electrons
is likely due to a different process than simple diffusion,
as discussed in the next section.  

The observation of hard synchrotron X-rays requires an efficient particle accelerator, 
since
in a transverse magnetic field the mean synchrotron photon energy ($E_{\gamma}$) scales as:
$$
E_{\gamma} \approx 2.2 (E_e/100 \, {\rm TeV})^2 (B/10 \, \mu {\rm G}) \, \, {\rm keV}
\eqno (2)
$$
Such energetic electrons cool rapidly, however, on a timescale of: 
$$
\tau_{sync} \approx 820 \, (E_e/100 \, {\rm TeV})^{-1} (B/10 \, \mu {\rm G})^{-2} \, \, {\rm year}
\eqno (3)
$$
Therefore the cooling timescale for electrons radiating photons $E_{\gamma}$ at 
mean energy is:
$$
\tau_{sync} \approx 1200 \, (E_{\gamma}/1 \, {\rm keV})^{-1/2} (B/10 \, \mu {\rm G})^{-3/2} \, \, {\rm year}
\eqno (4)
$$ 
Synchrotron cooling dominates over IC cooling for electron energies above 5 TeV
for the external photon and magnetic fields chosen, so we consider only synchrotron 
cooling in the following discussion. The electron cooling break results in a 
similar break in the emitted photon spectrum, which scales with
magnetic field and age ($\tau$) as:
$$
E_{\gamma,Break} \approx 26 \, (\tau / 13 \, {\rm kyr})^{-2} (B/10 \, \mu {\rm G})^{-3} \, {\rm eV}
\eqno (5)
$$
This break is apparent in the models plotted
in Figures 3 (break at $\sim 50$ eV) and 4 (break at $\sim$ 30 eV). 

The break location is also of key importance to IC modeling of the H.E.S.S. detection.  
Electrons upscatter blackbody photons (at temperature $T$) to a mean energy of: 
$$
E_{\gamma} \approx 2.2 \, (E_e/10 \, {\rm TeV})^{2}  \, (T/ 25 \, {\rm K}) \, \, {\rm TeV}
\eqno (6)
$$
where we scale $T$ to the value adopted for the far IR field.
HESS J1420--607 appears to peak in flux at $\approx 0.5-2.5$ TeV; we expect this peak to result
from a cooling break rather than an exponential cutoff in the electron spectrum
due to the need for significant numbers of VHE electrons to synchrotron radiate 
in the X-ray regime.  Inserting the value of $E_{\gamma} = 0.5-2.5$ TeV into Equation 6 
we see that at the break:
$$
E_{e,Break} \approx (5-11) \, (T/25 \, {\rm K})^{-1/2} \, \, {\rm TeV} 
\eqno (7)
$$
This VHE break is primarily due to synchrotron cooling, so substituting
Equation 7 into Equation 3 yields an estimate for the mean magnetic field strength in the nebula:
$$
B \approx (8-11) \, (T/25 \, {\rm K})^{-1/4} \, (\tau/ 13 \, {\rm kyr})^{-1/2} \, \, \mu {\rm G} 
\eqno (8)
$$
While this range of magnetic field (which ignores IC cooling) should not
be taken too seriously,
the fact that the 8 $\mu$G field we selected for single component modeling also matches the X-ray data 
lends some credence to this estimate.

Comparing the merits of each model, we find 
that the leptonic model provides a slightly better fit to the data, 
without the cost of the additional spectral components
of the hybrid model.   
In addition, in order for pion decay
to account for the VHE emission completely the energy
in protons exceeds $10^{51}$ erg
for a typical density of  $n=1 \, \rm cm^{-3}$, and the
fit to the lower energy H.E.S.S. data points is unimpressive.  
Even with $ > 10^{51}$ erg in hadrons, 
a primary electron population required to match the X-ray data contributes significantly to
the VHE flux via IC, as seen in Figure 4.  Only by raising the ambient magnetic field to 
 $> 20 \mu$G and lowering the total energy budget of this electron population can the IC from these
electrons be rendered insignificant.  Yet for a magnetic field of this magnitude synchrotron cooling
results in a 2--10 keV photon spectrum far steeper than our measured data, plus such a high 
magnetic field over such a large region would be surprising.   
The high energy requirement in hadrons, as well as the need for two, rather than one, highly tuned 
injection components makes the hybrid model a less appealing explanation for the observed extended
nebula
fluxes than a single electron population.  
In addition, further support for the leptonic model stems from its 
ability to account for the innermost
X-ray flux with a younger population of similar electrons radiating in the stronger magnetic field
near the pulsar.

Early detection by the Fermi LAT of the K3 nebula would undermine this assumption of a single
electron component, however, as in this model very little flux is expected in the GeV range.  Furthermore,
a photon index of $\alpha = (p+1)/2$ (where $p$ is the electron index) is expected, which
for our injection would imply a hard LAT index of 1.5.  Detection by the LAT is much more likely
in the hadronic scenario, and one would expect a GeV index of $\approx 2$.  
A LAT index steeper than 2 cannot be matched by either model, and 
would imply the existence of another particle population, likely electrons. 
Indeed, \citet{dj08} suggested the Vela X nebula could be modeled with two populations
of electrons.  Such a situation for the K3 nebula would allow for detection by the LAT, and might 
also explain the radio excess.  
LAT detection of the K3
nebula would also greatly aid in distinguishing between scenarios, though even LAT flux upper limits 
might preclude the hybrid model.

\subsection{Birthsite}

The radio, X-ray, and VHE data all hint at a pulsar birthsite to the north.  The possible radio
filament extending to the hotspot $8.5\arcmin$ to the North might mark the synchrotron trail
of J1420's proper motion.  
The nature of the X-ray (source 7) and radio point source 
at the end of this tail remains mysterious, though an AGN provides a likely candidate.  
The radio ``filament'' might also mark the eastern boundary of the apparent radio shell mentioned 
in Section 2.  Under this assumption, the pulsar could have originated $3 \arcmin$ 
northwest at the center of this shell.
With this assumption, electrons spewed off the pulsar and left behind by its motion should also
synchrotron radiate in the X-rays and emit VHE photons via IC scattering off field photons.
The observed X-ray tail to the North, and the H.E.S.S. position offset $\sim 3 \arcmin$ in this direction nicely
support this hypothesis.  
If the pulsar indeed originated $3\arcmin$ distant this implies a
reasonable velocity of
$$
v = 370 \, (\theta/3 \arcmin) \, d_{5.6} \, (t/ 13 \, {\rm kyr})^{-2} \rm \, km \, s^{-1}
\eqno (9)
$$ 
The northern tail could alternatively indicate an asymmetric reverse shock returning from the South
has crushed and displaced the K3 PWN, much like the situation observed in Vela X \citep{blondinetal01}.
The pulsar birthsite would remain unconstrained by our X-ray observations in this scenario.
Indeed, \citet{jandw06} measured the polarization angle of the polarization sweep, which they
argue correlates with the velocity.  The perpendicular of this line (marking the pulsar spin and 
likely proper motion axis if it is emitting in the orthogonal mode) points southwest back to the shell body
of the Kookaburra $\sim11 \arcmin$ away.  Association would require a very high space velocity of
$1400 \, d_{5.6} \, (t/ 13 \, {\rm kyr})^{-1} \rm \, km \, s^{-1}$ .

\section{Conclusions}

The Suzaku data provides the first deep mapping of the asymmetric extended X-ray emission
surrounding J1420.  The spectral index appears to soften in the outer nebula, as expected
in a leptonic model with synchrotron cooling of electrons.  SED studies of this nebula allow for
a hybrid leptonic $+$ hadronic model, yet favor a pure leptonic model if one takes into account energy
requirements and the degree of tuning required to fit the hybrid model.  
The outer nebula also yields information about the pulsar birthsite, hinting at 
an origin to to the North, though admittedly the northern X-ray tail could be due to reverse shock
interaction with the PWN and therefore not correspond to the pulsar velocity.  
Further investigation of the apparent radio shell along with deeper Chandra imaging of the compact nebula
might also help elucidate the origin of PSR J1420--6048.  

{\it Acknowledgments: We thank Stefan Funk for useful comments on this paper.}

\end{document}